\documentclass[12pt]{article}
\usepackage{amssymb,amsmath}
\usepackage{graphicx}
\usepackage{ulem}
\begin{document} 
\begin{center} 
{\bf\Large {Fusion of halo nucleus $^{6}He$ on $^{238}U$: evidence for tennis-ball (bubble) structure
of the core of the halo (even the giant-halo) nucleus}} 
\end{center} 
\vspace{.1in} 
\begin{center} 
{\bf Syed Afsar Abbas}\\ 
\vspace{.05in} 
Centre for Theoretical Physics, JMI University, New Delhi-110025, India\\
and\\
Jafar Sadiq Research Institute, T-1 AzimGreenHome, NewSirSyedNagar, Aligarh - 202002, India\\
email:  drafsarabbas@gmail.com
\end{center} 
\begin{center} 
{\bf Abstract} 
\end{center} 

In a decade-and-a-half old experiment, Raabe et al.(Nature 431 (2004) 823), had studied fusion of an incoming beam of halo nucleus $^{6}He$ with the target nucleus $^{238}U$. We extract a new interpretation of the experiment, different from the one that has been inferred so far. We show that their experiment is actually able to discriminate
between the structures of the target nucleus (behaving as standard nucleus with density distribution described with canonical RMS radius r = $r_0 A^{\frac{1}{3}}$ with $r_0$ = 1.2 fm), 
and the {\bf "core"} of the halo nucleus, which 
surprisingly, does not follow the standard density distribution with the above RMS radius.
In fact the core has the structure of a tennis-ball (bubble) like nucleus, with a "hole" at the centre
of the density distribution. This novel interpretation of the fusion experiment provides an unambigous support
to an almost two decades old model (Abbas, Mod. Phys. Lett. A 16 (2001) 755), of the halo nuclei.
This Quantum Chromodyanamics based model,
succeeds in identifyng all known halo nuclei and makes clear-cut and unique predictions for 
new and heavier halo nuclei. This model supports the existence of tennis-ball (bubble) like core,
of even the giant-neutron halo nuclei.
This should prove beneficial to the experimentalists, to go forward
more confidently, in their study of exotic nuclei.


\newpage

Right from the first completely  unexpected appearance of the two-neutron halo stucture in $^{11}Li$
in 1985, the exotic nuclei have been focus of intense interest in  nuclear physics. 
Hence, really not unexpectedly, the theoretical predictions of new halo nuclei, and their subsequent experimental searches, 
have had a tortuous history.
Suffice to quote the example of $^{24}O$ (with Z=8, N=16), predicted and expected to have a neutron halo structure,
was eventually and again shockingly, found to be actually a strongly doubly-magic nucleus!
Clearly there is something amiss in our theoretical understanding of the nuclear halo phernomenon.
What is it? Here we show that the answer lies in an internally consistent study of a decade-and-a-half
old nuclear fusion experiment.

Within the framework of the studies of neutron rich nuclei, it is of great importance to know whether
fusion of nuclei involving weakly bound particles  is enhanced or not.
We look at the experimental data in this connection, and try to extract some basic structures which these
ingenious experiments are trying to point out to us.

Raabe et al. [1] fired both $^{4}He$ and 2-neutron halo nucleus $^{6}He$ onto the target nucleus $^{238}U$.
In agreement with an earlier experiment [2], they did obtain a much increased product with the above neutron halo beam. These large yields due to fission, may be attributed to fusion of $^{6}He$ on the target-nucleus. However, the same may be due to a transfer of neutrons from $^{6}He$, first onto $^{238}U$, and thereafter a fission from this fattened nucleus. The earlier experiment [2]
was unable to distinguish between these two possibilities. 
Raabe et al. [1] cleverly, were able to distinguish between the above two physically possible
occurrences. If a fission event was detected in coincidence with an $^{4}He$, it was identified with the transfer case, while 
one without an accompanying $^{4}He$, was attributed to complete fusion.
Remarkably, they demonstrated clearly, that the large fission yields do not result from fusion with $^{6}He$, but from neutron transfer. Thus the {\bf "core"} of the projectile nucleus sees a larger nucleus which has "eaten and digested" the halo neutrons [3].
So what is amazing here, is the new phenomenon of fusion only of the projectile "core" with a neutron-fattened target nucleus, which may itself be in an excited state [3].

What is this experiment trying to tell us? Using Ockham's razor, what it is telling us is that though the 2-neutron halo is
weakly bound with the {\bf core} $^{4}He$ in $^{6}He$, it is strongly attracted to the target nucleus.
Hence minimal requirement is that the {\bf "core"} of the halo,  and the {\bf "target"} itself, should differ from each other, is some minimal significant manner.
Now we know that the density distribution of the standard nuclear medium is given  by the RMS radius 
r = $r_0 A^{\frac{1}{3}}$ with $r_0$ = 1.2 fm).
This is definitely true of the target nucleus $^{238}U$. And as the two neutrons (from the projectile nucleus)
feel strong nuclear attraction with it, we would expect that the neutron fattened target nucleus would be a
standard (though perhaps excited) nucleus with density distribution conforming to the above standard nuclear RMS radius.
This means that, therefore, the {\bf "core" of the projectile nucleus should be different from the initial target nucleus}, in
some fundamental manner. Note how this fusion experiment allows us to talk of the density distribution of the "core"
of the whole halo nucleus.
The beauty of this experimental analysis by Raabe et al. [1], is that it is allowing us to 
separate out the structure of the core-nucleus itself, which is 
sitting within the whole halo nucleus.
But what does it mean?

\begin{figure}
\vspace{0.50cm}
\includegraphics[scale=0.80]{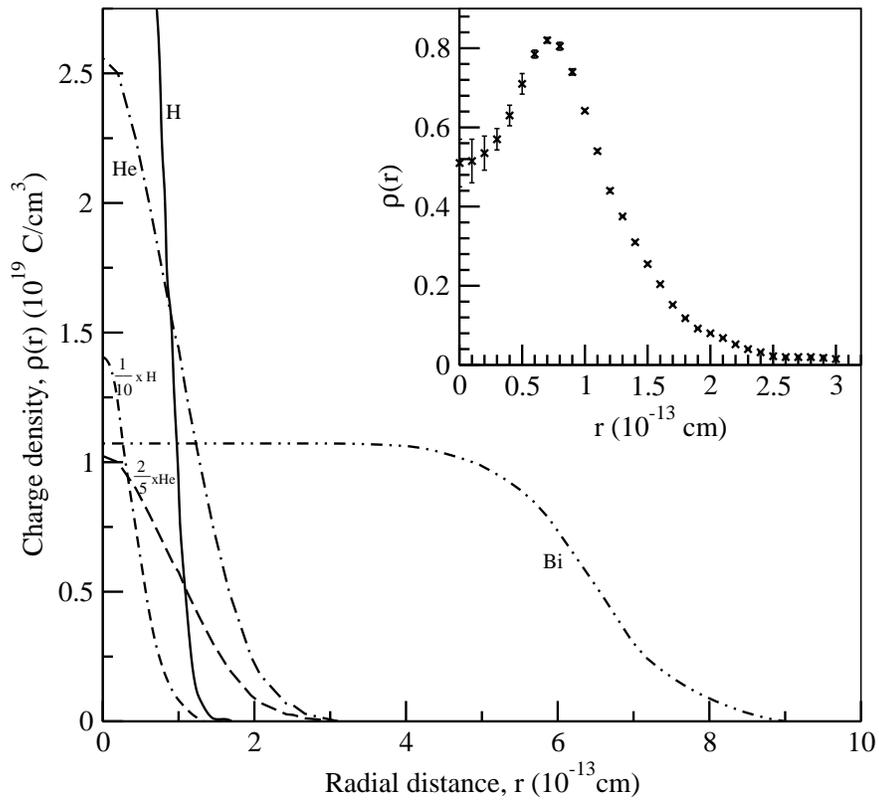}
\caption{  Schematic density distributon of nuclei as determined by electron scattering. Inset shows the same
from a better experiment - showing a marked "hole" at the centre.}
\label{Fig. 1}
\end{figure}

I  would like to draw the reader's attention to the density distribution as was extracted from the classical electron scattering on nuclei.
This is plotted in Fig. 1 here. These are well known figures.
What is most significant is that the central density of $^{4}He$ is about 2.5 times higher than that of heavy
nuclei like bismuth, lead etc.. In fact, $^{3}He$ has similar density distribution as that of $^{4}He$. Not only  that,
as determined from meticulous electron scattering experiemts, both $^{4}He$ and $^{3}He$ have a "hole" 
(the central significant depression near $r \rightarrow 0$) at the centre. This is plotted in the inset of 
Fig. 1 [4,5].
It is known that light nuclei are basically "all surface". Because of the central hole,  this is more pronounced for these two 
nuclei. Note that as to matter distribution of $^{3}H$, it is very much the same as that of $^{3}He$.

Now to understand the fusion process of neutron halo beam, forced us to conclude logically, using  the Ockham's Razor based
argument, that 
the core of the projectile nucleus should be different from the target nucleus in
some fundamental manner. This can now be extracted from a study of Fig. 1. The much-bigger-in-magnitude
"surface-nature" of density 
distribution of $^{4}He$ plus its hole at the centre, should be the reason of this "fundamental-manner-difference", between the target and the core of the projectile nucleus.

Thus the density distribution of the core of the halo nucleus, here $^{6}He$, but in general any neutron halo nucleus,
has a tennis-ball (bubble) like structure. This is the
novel interpretation extracted from this halo fusion experiment, and which has not been understood in this manner, as of now.
This new interpretation provides an unambiguous suppport to an almost two-decades-old
Quantum Chromodynamics based model predictions by the author [6].

\begin{figure}
\vspace{0.50cm}
\includegraphics[scale=0.50]{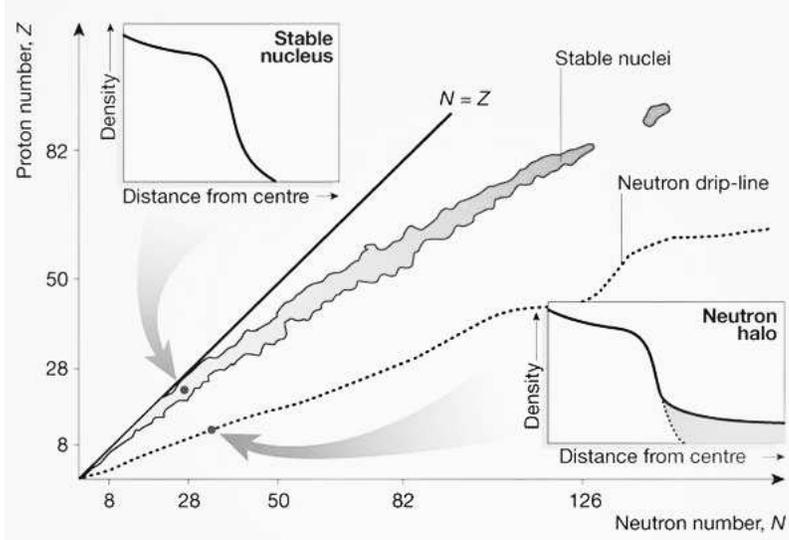}
\caption{  Schematically drawn from Fig. 1 of Ref. [3]}
\label{Fig. 0}
\end{figure}

But what has been the standard interpretation of the density distribution of the core of the halo nucleus
in the above fusion experiment of $^{6}He$ on $^{238}U$? 
This is best explained by looking at Fig. 1 of ref. [3], which we display here schematically as Fig. 2. 
The density distribution on the left-inset, as that the stable nucleus, is clearly that of the target nucleus $^{238}U$. The right-inset shows density distribution of the halo nucleus, here 
$^{6}He$, but in general of any other putative halo incoming beam. Note the identical density distribution structure of the core of the halo nucleus with that of the left inset figure. The only difference there, is due to the additional extension of the halo neutrons. The same point is also emphasized by Tanihata [7], "The density distribution of the core 
is assumed to be the same as the bare ground state nucleus". {\bf Above, we saw that this assumption is actually quite wrong}.
However, this is the picture of the halo nuclear density distribution, which ubiquitously pervades all the theoretical 
model structures available at present; {\bf except}, of course, the 2001 model as pointed above [6].

Arguments originating from  Quantum Chromodynamics, have allowed us to 
provide an understanding of the  unusual phenomenon of the halo structure as part of the overall nuclear phenomenon.
The author thus arrived at a model which could explain the existence of all known halo nuclei, and provide clearcut predictions 
for many more halo nuclei - which were actually discovered later, and thus validating this model [6].
For the sake of better comprehension, now onto a brief exposition of that model [6].

As per Quantum Chromodynamics, the physically observed hadrons correspond to
the colour singlet representation. So for a baryon
in $3\otimes3\otimes3 = 1 \oplus 8 \oplus 8 \oplus10$, of all the representations, it is only the singlet
which provides observable baryons. All the other colour representations seem to be spurious and unnecessary.
But this may not be true for multiquark systems, where $8\otimes8 = 1 + ....$ and hence in 6-quarks this colour singlet representation
may be present [8]. But in nuclear physics, we treat the sysytem as made up only of individual colour singlet protons and neutrons, with the commonly
held belief that no quarks would show up in low energy nuclear physics; and only at sufficiently high energies,
these may manifest themselves in terms of a Quark-Gluon-Plasma. However, this naive view is not correct.
We show here, that even at low energies, quarks do place their identifiable imprints in a nucleus.

Deuterons should have 
configurations where the two
nucleons overlap strongly 
in regions of size $ \leq 1 fm $ to form 6-quark bags. Why is
deuteron such a big and loose system? The reason has to do with the
structure of the 6-q bags formed, had the two 
nucleons overlapped
strongly.
As per the colour confinement hypothesis of QCD, the 6-q wave function looks like:

\begin{equation}
| 6q > = \frac{1}{ \sqrt{5} } | 1 \otimes1 > + \frac{2}{ \sqrt{5} } | 8 \otimes 8 >
\end{equation}
where '1' represents a 3-quark cluster which is singlet in colour space 
and '8' represents the same as octet in colour space. Hence $ | 8 \otimes 8 > $ 
is overall colour singlet. This part is called the hidden colour 
because as per confinement ideas of QCD,
these octets cannot be separated out asymptotically, and so manifest
themselves only within the 6-q colour-singlet system. 
Group theoretically the above hidden colour part of $ 80 \% $ was determined by Matveev and Sorba [9].
This large hidden colour part would prevent the two nucleons to come together and overlap
strongly [8,9]. Therefore the hidden colour would manifest itself as
short range repulsion in the region $ \leq 1 fm $ in deuteron. So the two
nucleons though bound, stay considerably away from each other.

For the ground state and 
low energy description of nucleons, we 
assume that the group $ SU(2)_{F} $, with u- and d-quarks in the fundamental trepresentation, is 
what is required. Hence we
assume that 9- and 12-quarks belong to the totally antisymmetric
representation of the group 
$ SU(12) \supset SU(4)_{SF} \otimes SU(3)_{C} $ where
$ SU(3)_{c} $ is the QCD group and  
$ SU(4)_{SF} \supset SU(2)_{F} \otimes SU(2)_{S} $  where S denotes spin.
Note that up to 12-quarks can sit in the s-state in the group SU(12) [8]. The
calculation of the hidden colour components for 9- and 12-quark systems
requires the determination of the coefficients of fractional parentage 
for the group $ SU(12) \supset SU(4) \otimes SU(3) $,
which becomes quite complicated for a large number of quarks [10]. 
The relevant group theoretical techniques were developed by So and Strottman [11] and independently by Chen [12],
Thereafter, the  author, group theoretically determined [8,13] that the hidden colour component of the 9-q
system is $ 97.6 \% $ while the 12-q
system is $ 99.8 \% $ i.e. almost completely coloured.

What is the relevance of
these 9- and 12-quark configurations in nuclear
physics? The A=3,4 nuclei 
$ ^{3}H $, $ ^{3}He $ and $ ^{4}He $ have sizes of 
1.7 fm, 1.88 fm and 1.674 fm respectively. Given the fact that each
nucleon is itself a rather diffuse object, quite clearly in a size
$ \leq 1 fm $ at the centre of these nuclei, the 3 or 4 nucleons would
overlap strongly. As the corresponding 9- and 12-q are predominantly
hidden colour, there would be an effective repulsion at the centre keeping
the 3 or 4 nucleons away from the centre. Hence it was predicted by the
author [13] that there should be a hole at the centre of 
$ ^{3}H $, $ ^{3}He $ and $ ^{4}He $.
And indeed, this is what is found through electron scattering [4,5]. 
This is shown as inset in Fig. 1 here. Hence the hole, i.e. significant depression in the
central density of $ ^{3}H $, $ ^{3}He $ and $ ^{4}He $, is a signature of quarks in this
ground state property.

This understanding of the hole, within the above QCD based arguments, leads us further to
provide a consistent understanding of the halo structure pheneomenon as well [6].

Due to the significantly higher
density at the boundary and very small at the centre, 
$ ^{4} He $ is like a "tennis-ball". The word tennis-ball is used to emphasize the 
predominance of the
"surface-ness" property in the density distribution in the corresponding nuclei.
Add two more neutrons to 
$ ^{4}He $ to make it $ ^{6}He $, a bound system. As the two neutrons approach the
surface, they will bounce off. As the two neutrons are bound, these
will ricochet on the compact tennis-ball like nucleus. 
A neutron halo would be manifested as these neutrons
shall be kept significantly away from the core.

How do we understand this effect?
Macroscopically, we understand this as: the density of the $ ^{4}He $ core is high on the
boundary, any extra neutrons would not be able to penetrate it, as this would
entail much larger density on $ ^{4}He $ surface than the system would
allow dynamically. Microscopically, we understand this as: any penetration of extra neutron
through the surface of $ ^{4}He $ would necessarily imply the existence of
five or six nucleons at the centre. As already indicated, due to the
relevant SU(12) group, only 12-quarks can sit in the s-state, which already
is predominantly hidden colour. Any extra quarks hence would have to go to
the p-orbital; and in the ground state of nucleus, there is not sufficient
energy to allow this. Hence, the two bound neutrons are consigned to stay outside
the $ ^{4}He $ boundary. In addition, if at any instant the two neutrons
come close to each other while still being close to the surface, locally
the system would be like three nucleons overlapping, and which would look like a 9-q
system. This too would be prevented due to the local hidden colour repulsion.
Hence as found experimentally, the two neutrons in the halo would not come
close to each other, resulting in the neutron halo in $ ^{6}He $ [6].

  Going through the binding energy systematics of neutron rich nuclei, one
notices that as the number of 
$ \alpha $'s increases along with the neutrons, each $ ^{4}He $ + 2n pair
tends to behave like a cluster of two 
$ ^{3}_{1}H_{2} $ nuclei. Remember that though 
$ ^{3}_{1}H_{2} $ is somewhat less strongly bound (ie. 8.48 MeV ), it is
still very compact (ie. 1.7 fm ), almost as compact as $ ^{4}He $ (1.674
fm). In addition it too has a hole at the centre. Hence $ ^{3}H $ is also a
tennis-ball like nucleus. 

Hence $ ^{7}Li $ which is 
$ ^{4}He + ^{3}H $; with two more neutrons it becomes $ ^{9}Li $, which can be treated as made
up of  $3\; ^{3}_{1}H_{2} $ clusters. 
Let us treat tritons as sitting at the vertices of an equilateral triangle. Because of the
tennis-ball like structure, the $3\; ^{3}_{1}H_{2} $ particles cannot come too
close to each other. Firstly, the surface of the ball would prevent it from doing so, and
secondly if some part of the $3\; ^{3}_{1}H_{2} $ still overlap at the centre,
it would look like a 6- or 9-quark system. Therein the hidden colour
components would repel, ensuring that the $3\; ^{3}_{1}H_{2} $ clusters do not
approach too closely at the centre. This too would imply a depression in
the central density of $ ^{9}Li $. 
$ ^{12}Be $ treated as $4\; ^{3}_{1}H_{2} $ sitting at the vertices of 
a regular tetrahedron would, for the
reasons stated above, too have a central density depression. Thus
$ ^{9}Li $ and $ ^{12}Be $ would appear more surface-like or tennis-ball-like as well.
Other evidences like the actual decrease of radius as one goes from
$ ^{11}Be $ to $ ^{12}Be $ [7, see Fig. 4] supports the view that it ( i.e. $ ^{12}Be $ )
must be made up of four compact clusters of $ ^{3}H $.

Next, the tennis-ball like nature of $ ^{3}H $ and $ ^{3}He $ has a unique structural property which even
$ ^{4}He $ does not have. The nuclei $ ^{3}H $ and $ ^{3}He $ along with deuteron, are the only known nuclei
which have no excited state. Either these nuclei are there, or are not there, as a single rigid entity. Due to quantum mechanics,
right upto their binding energy of 8.48 MeV, tritons would be immune to any excitations; and thus their
rigidity/elasticity of a tennis-ball
like nature, would be more explicitly exhibited.

What we are saying is, that
the neutron rich nuclei, which are made up of a number of
tritons, each of which is tennis-ball like and compact, should be compact
as well. These too would develop tennis-ball like property. This is,
because the surface is itself made up of tennis-ball like clusters. 
Hence when
more neutrons are added to this ball of triton clusters, these 
extra neutrons will ricochet on the surface.
Hence we expect that one or two or more bound neutrons outside these compact clusters,
would behave like neutron halos. Let $ ^{9}Li $ be treated as made
up of $ 3 ~^{3}H $ clusters and which should have hole at the centre. 
Therefore $ ^{11}Li $ with $^{9}Li + 2n $
should be two neutron halo nuclei - which it is. So should 
$ ^{14}Be $ be. It turns out that internal dynamics of $ ^{11}Be $
is such that it is a cluster of $ \alpha - t - t $,
with one extra neutron halo around it [6].
Similarly e.g.  $ ^{17}B, \;^{19}C, \;etc.$
would be neutron halo nuclei and so on. These specific predictions of 2001
have been confirmed in later years [7,14]. 

Still more heavy nucleus $ ^{31}Ne $, predicted to be one neutron-halo as per our model,
found experimental confirmation in 2009 [14]. Still heavier nucleus $^{37}Mg $, was found to be a one
neutron halo nucleus in 2014 [15], one more confirmation of our model.
All this provides clear and unambiguos support to our model. 
$^{37}Mg $ remains the heaviest halo nucleus discovered so far.
However our model predicts many more heavier halo nuclei; and the experimentalists are urged to look for those.

The proton halo nuclei can also be understood in the same
manner. Here, another nucleus with a hole at the centre 
$ ^{3}_{2}He_{1} $ (binding energy 7.7 MeV, size 1.88 fm), would play a
significant role.

Thus all light neutron rich nuclei $ _{Z}^{3Z}A_{2Z} $ are made up of Z 
$ ^{3}_{1}H_{2} $ clusters. 
Recently we used the RMF model with the best interactions, to study the predictions of this model [16].
That [16] too has confirmed our model [6].

Note that we prefer to use the word tennis-ball, vis-a-vis the word bubble.
Firstly, because while the word bubble, connotes ephemeral nature of the entity, the word
tennis-ball signifies its rigidity/elasticity/stability.
Next, the surface of the tennis-ball is more resilient, so that a neuton coming from outside, 
can at best, just ricochet on top. Thus the stabilty of the core, and the halo nature of the extra neutrons,
is connoted more naturally in it. However we use both the words here, due to the fact that in the nuclear physics literature,
it is more common to come across the word bubble, rather than the word tennis-ball.
However, in connection with the core of the
halo, it is clearly better to use the word, tennis-ball like, rather than bubble like. 

This brings into focus the recently discovered [17] bubble nature of the nucleus $ ^{34}Si$. Our model predicts that 
$ ^{42}Si = 14 ^{3}H$, be a bound system with a hole at the centre. Hence the bubble nature of $ ^{34}Si$, 
as observed by them [17], should be just 
a remnant effect of the more strongly bound tennis-ball like state of $ ^{42}Si$.
We urge the expeimentalists to look for it.
 
We have treated all  neutron rich nuclei $ _{Z}^{3Z}A_{2Z} $, as made up of Z number of 
$ ^{3}_{1}H_{2} $ clusters. This is easily understood geometrically, as say 3-tritons sitting at the vertices of an equilateral traingle for $ _{3}^{9}{li}_{6} $,
and 4-tritons sitting at the vertices of a regular tetrahedron for $ _{4}^{12}{Be}_{8} $.
In Table 1, we show several neutron rich nuclei which may be treated 
 as being composed of n-clusters of  $^3_1 \rm H_2$. 
 We write the binding energy of these nuclei as [6],
 \begin{equation}
 E_B = 8.48 n + C k 
 \end{equation}
where 8.48 MeV is the binding energy of $^3_1\rm H_2$, with
n-cluster of tritons forming k bonds and with C being
 inter-triton-bond energy.
 Here we take the same geometric structure of clusters in 
 these nuclei as conventionally done for $\alpha$ - clusters in A=4n nuclei. 
 Thus this model seems to hold out well with inter-triton cluster bond energy 
 of about 5.4 MeV, which continues to work for even heavier neutron rich nuclei, 
  e.g. for $_{14}^{42}{\rm Si}_{28}$,

 \begin{table}[ht]
\caption{Neutron-rich nuclei - inter-triton cluster bond energy}
 \renewcommand{\tabcolsep}{0.40cm}
 \begin{tabular}{ccccc}
 \hline\hline
 Nucleus& n& k& ${\rm E}_{\rm B}$-8.48n(MeV) &C(MeV)\\
 \hline
 $^{9}{\rm Li}$ &3&3 &19.90 &6.63\\
 $^{12}{\rm Be}$&4&6 &34.73 &5.79\\
 $^{15}{\rm B}$ &5&9 &45.79 &5.09\\
 $^{18}{\rm C}$ &6&12&64.78 &5.40\\
 $^{21}{\rm N}$ &7&16&79.43 &4.96\\
 $^{24}{\rm O}$ &8&19&100.64&5.30\\
 \hline\hline
 \end{tabular}
 \label{tab1}
 \end{table}

 Hence the tennis-ball (bubble) like structure seems to hold good for even heavy nuclei.
 We can understand this as per QCD as follows. 
As already indicated, due to the
 relevant SU(12) group, only 12-quarks can sit in the s-state, which already
 is predominantly hidden colour. 
 Hence if two tritons come close to each other in a heavy triton-rich nucleus, then it would necessarily imply
 overlap of six nucleons at the centre of mass of these overlapping tritons.
 Any extra quarks above the number twelve, relevant for the group SU(12), hence would have to go to
 the p-orbital; and in the ground state of that nucleus, there is not sufficient energy to allow this. 
 Hence, any two bound tritons are consigned to stay away from each other.
 Hence, this would result in the tennis-ball (bubble) like structure in say, $_{14}^{42}{\rm Si}_{28}$  [6].
 
 However, in this context, interesting is the concept of "giant-halo-nucleus", as proposed by Meng and Ring [18] for
 halo in Zr nuclei (Z=40, $A > 82$), 
 and later by Zhang, Meng, Zhou and Zeng [19] for halo in experimentally more accessible region in Ca-isotopes with
 $A > 60$. Note that our model fundamentally agrees with this prediction of giant-halo-nuclei.
 Thus e.g. as per our model, $^{124}Zr \sim ^{120}Zr + 4n = 40 ^{3}H + 4n$, and
 $^{64}Ca \sim ^{60}Ca + 4n = 20 ^{3}H + 4n$. Thus  $^{124}Zr$ and $^{64}Ca$ 
 should both be four-neutron halo nuclei. Similarly for other heavier Zr and Ca isotopes.
It is amazing that these giant-halo nuclei, are clearly demanding tennis-ball like core of such
 heavy nuclei as  $^{120}Zr$ and $^{60}Ca$.
 
 Recently stabilty of $ ^{60}Ca $ has been demonstrated empirically [19].
 This is yet another confirmation of our model.
 The same experiment [19] also discovered several new neutron-rich nuclei as,
 $ ^{47}P, \; ^{49}S, \; ^{52}Cl, \; ^{54}Ar, \; ^{57}K,\; ^{59}K, \;  ^{62}Sc$.
 These find clear interpretation in our model. For example,  $ ^{47}P = ^{45}P + 2n = 15 ^{3}H + 2n$, and hence 
 is a 2-neutron halo ouside a core of $ ^{45}P = 15 ^{3}H$ - a tennis-ball like and compact core of the halo nucleus.

The successful application of the structures arising from
proper interpretation of the fusion probability with $^{238}U$ 
by the neutron halo of $^{6}He$, leads to unambiguos support to our 
QCD based model. At the base sits the prediction of a prominent "surface-structure" in the density distribution 
along with a hole at the centre of the core of all the halo nuclei. 
However, the recent work of experimentally determining the density
distribution of halo nuclei with electron scattering (as emphasized by Bertulani [20]) is very exciting.
We look forward to advanced precisions in these experiments, which will allow them to see the central density depression 
{or a "hole"} in these neutron-rich nuclei, as uniquely predicted by our model.

\newpage
{\bf REFERENCES}


1. R. Raabe, J. L. Sida, J. L. Charvet, N. Alamanos, C. Angulo, J. M. Casandjian,
   S. Courtin, A. Drouart, D. J. C. Durand, P. Figuera, A Gillibert, S Heinrich, C Jouanne, V Lapoux,
   A Lepine-Szily, A Musumarra, L Nalpas, D Pierroutsakou, M Romoli, K Rusek and M Trotta, Nature 431 (2004) 823

2. M. Trotta et al., Phys. Rev. Lett. 84 (2000)  2342 

3. D. Hinde and M. Dasgupta, Nature 431 (2004) 748 

4. R. G. Arnold et al., Phys. Rev. Lett. 40 (1978) 1429 

5. I. Sick, Lecture Notes in Physics,Vol.87(Sptringer,Berlin, 1978)p. 236   

6. A. Abbas, Mod. Phys. Lett. A 16 (2001) 755 

7. I. Tanihata, Nucl. Phys. A 685 (2001) 80c

8. S. A. Abbas, "Group Theory in Particle, Nuclear, 
and Hadron Physics", CRC Press, Boca Raton, Florida, 2016 

9. V. A. Matveev and P. Sorba, Lett. Nuovo Cim., 45 (1978) 257

10. M. Harvey, Nucl. Phys, A 352 (1981) 301, 326 

11. S. I. So and D. Strottman, J. Math. Phys. 20 (1979) 153

12. J. Q. Chen, J. Math. Phys. 22 (1981) 1 

13. A. Abbas, Phys. Lett. B 167 (1986) 150; Prog. Part. Nucl. Phys., 20 (1988) 181 

14. I. Tanihata, H. Savajols, R. Kanungo , Prog. Part. Nucl. Phys. 68 (2013) 215

15. N. Kobayashi et al., Phys. Rev. Lett. 112 (2014) 242501

16. A. A. Usmani, S. A. Abbas, U. Rahaman,
M. Ikram and F. H. Bhat, Int. J. Mod. Phys. E 27 (2018) 1850060

17. A. Mutschler, A. Lemasson, O. Sorlin, D. Bazin, C. Borcea, R. Borcea, Z. Dombradi,
J.-P. Ebran, A. Gade, H. Iwasaki, E Khan, A Lepailleur, F Recchia, T Roger, F Rotaru, D Sohler,
M Stanoiu, S R Stroberg, J A Tostevin, M Vandebrouck, D Weisshaar and K Wimmer, 
Nature Physics 13 (2017) 152

18. 7. J. Meng, P. Ring, Phys. Rev. Lett. 80 (1998) 460;
S.-Q. Zhang, J Meng, S.-G. Zhou, J.-Y. Zeng, Chi. Phys. Lett. 19
(2002) 312

19. O. B. Tarasov,D S Ahn,D Bazin,1 N Fukuda,A Gade,M Hausmann,N Inabe,S Ishikawa,
N Iwasa,K Kawata,T Komatsubara,T Kubo,K Kusaka,D J Morrissey,M Ohtake,H Otsu, 
M Portillo,T Sakakibara,H Sakurai,H Sato,B M Sherrill,Y Shimizu,A Stolz,T Sumikama, 
H Suzuki,H Takeda,M Thoennessen,H Ueno,Y Yanagisawa and K Yoshida,
Phys Rev Lett 121(2018)022501

20. C. A. Bertulani, J. Phys. G 34 (2007) 315

\end{document}